# High flux lithium antineutrino source with variable hard spectrum. How to decrease the errors of the total spectrum ?

**V. I. Lyashuk**[1,2]

[1]Institute for Nuclear Research, Russian Academy of Sciences, 117312 Moscow, Russia
[2]National Research Center Kurchatov Institute, 123182, Moscow, Russia

The proposed high flux lithium antineutrino source is based on $^{7}Li(n,\gamma)^{8}Li$-activation and subsequent fast $\beta^{-}$-decay ($T_{1/2}$ = 0.84 s) of the $^{8}Li$ isotope with emission of hard $\tilde{\nu}_e$ with energy up to 13 MeV. In the discussed scheme the nuclear reactor is used as an intensive neutron source for (n,γ)-activation. The created $^{8}Li$ isotopes are transported in the channels with a high speed to the remote neutrino detector and removed back for the next activation. The antineutrino source operates in the continuous cycle in the closed loop. Use of $^{8}Li$ hard $\tilde{\nu}_e$-spectrum in the source scheme ensure significantly harder total antineutrino spectrum (compare to the reactor one) in the position of the detector. The analyses of errors of the total $\tilde{\nu}_e$-spectrum (from reactor antineutrinos plus from lithium ones) confirmed that owing to the lithium $\tilde{\nu}_e$-spectrum the errors of the total $\tilde{\nu}_e$-spectrum can be diminished in two times and more. The results of errors analysis is important for sterile neutrino search and the possible configuration of the experiment is discussed.

## 1. INTRODUCTION. LITHIUM ANTINEUTRINO SOURCE WITH VARIABLE AND REGULATED TOTAL $\tilde{\nu}_e$-SPECTRUM

This article is the next step in development of the high flux lithium antineutrino source with variable spectrum and really is the continue of the work [1], devoted to the main principles for creation of the lithium $\tilde{\nu}_e$-source with transport of the activated $^{8}Li$ to the position of the neutrino detector.

It is necessary to emphasize the main items which important for the purpose of this article. The principles which were based for the antineutrino source are clear from the Fig. 1. Liquid lithium (or heavy water solution of $^{7}Li$ chemical compound) is pumped over in a closed cycle through a lithium blanket (placed around the AZ - active zone of the reactor) and further in a direction to a remote neutrino detector. The regime of pumping (including the control of all parameters) is supported by the installation for maintenance of the regime (not shown in the Fig. 1) In order to increase the part of hard $^{8}Li$ antineutrinos a being pumped volume (volumable reservoir) is placed near the $\tilde{\nu}_e$-detector.

Due to the geometrical factor the total $\tilde{\nu}_e$-spectrum in the detector volume (i. e., the resulting $\tilde{\nu}_e$-spectrum formed by $\tilde{\nu}_e$-flux from the reactor active zone plus from $\beta^{-}$-decay of $^{8}Li$) will be more harder compare to reactor antineutrino spectrum. It is clear that the closer to the reservoir will be detector the total spectrum will be harder. From the other side if the detector will be farther from the reactor then the part of the soft reactor $\tilde{\nu}_e$-spectrum in the total spectrum will be more small. For the fixed distance $L_1$ between the lithium blanket and the reservoir and $L_2$ (from the detector to the reservoir) the most harder spectrum is ensured for the position along the direction $C$-$C'$.

Such type of a facility will ensure not only more hard total spectrum in the location of a detector but also an opportunity to change smoothly the hardness of the spectrum in the simple way varying a rate of lithium pumping in the channels. So, this source is the installation with regulated and controlled total $\tilde{\nu}_e$-spectrum.

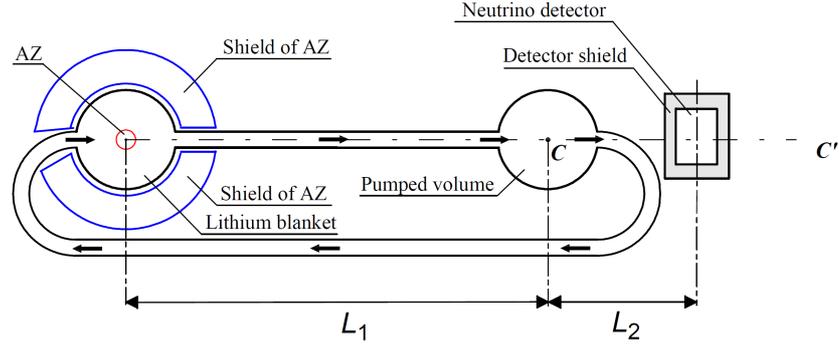

Fig. 1. Scheme of the lithium antineutrino source with variable (regulated and controlled) total $\tilde{v}_e$ - hard spectrum.

In the discussing of the spectrum errors we will base on the definition of the hardness (or generalized hardness) of the total $\tilde{v}_e$-spectrum [1]. The value of the hardness $H$ is defined the relation between lithium $F_{Li}(\vec{r})$ and reactor densities $F_{AZ}(\vec{r})$ of $\tilde{v}_e$-fluxes at the point $\vec{r}$:

$$H(\vec{r}) = \bar{n}_v \frac{F_{Li}(\vec{r})}{F_{AZ}(\vec{r})} , \qquad (1)$$

where $\bar{n}_v$ = 6.14 - number of reactor antineutrinos emitted per one fission in the active zone.

Fig. 2 illustrate change of the total $\tilde{v}_e$-spectrum for different hardness of the resulting antineutrino spectrum (in case of the one fuel isotope - $^{235}$U).

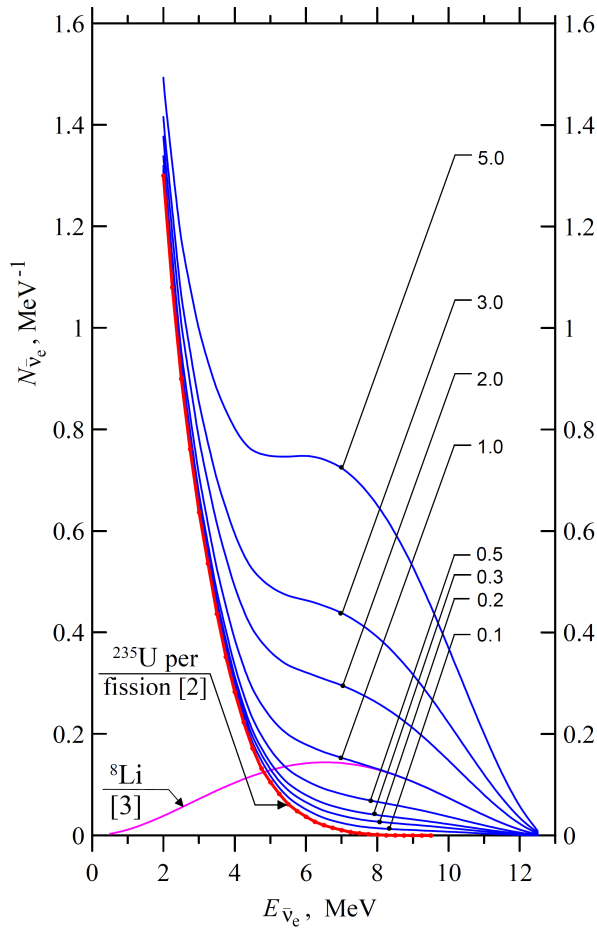

Fig. 2. Total $\tilde{v}_e$-spectrum [blue lines] (produced by $\tilde{v}_e$-flux from decays of fission products of $^{235}$U [red colour] and $\tilde{v}_e$ from $^8$Li [violet]) corresponding to different it's hardness $H$.

Increasing part of hard lithium antineutrinos in the total $\tilde{\nu}_e$-spectrum means rise of the hardness $H$ of the resulting spectrum. The hardness value can be increased by the: 1) re-configuration of the scheme geometry (position of the detector relative to the blanket and active zone; distance between active zone and pumped volume); 2) increase of the rate of pumping (volume pumped through the channels per unit of time); 3) use of the blanket with higher productivity (efficiency) - number of $^8$Li produced in the blanket volume per fission in the active zone. So, the possibility to regulate and control the spectrum means to vary the it's hardness $H$. The hardness $H$ can be varied smoothly by varying the rate of lithium pumping.

## 2. ERRORS OF THE TOTAL $\tilde{\nu}_e$-SPECTRUM FOR INCREASING GENERALIZED HARDNESS. HOW TO DECREASE THE ERRORS ?

In the [1,4] we note that $\tilde{\nu}_e$-cross section in the total spectrum is the additive value of the cross sections in the $\tilde{\nu}_e$-flux from the active zone and from $^8$Li antineutrinos. In fact the total number of $\bar{\nu}_e$ (entering to the neutrino detector) is:

$$N_{\tilde{\nu}_e} = N_{AZ} + H(\vec{r})\frac{N_{AZ}}{\bar{n}}, \qquad (2)$$

where $N_{AZ}$ - number of $\bar{\nu}_e$ from the active zone, $\bar{n}_\nu$ - number of $\bar{\nu}_e$ from the active zone per one fission, $H(\vec{r})$ - averaged generalized hardness of the total spectrum in the detector position. The second summand determines the number of lithium antineutrinos.

More strictly for density of the total $\tilde{\nu}_e$-flux in the point $\vec{r}$ we can write:

$$F_{\tilde{\nu}_e}(\vec{r}) = F_{AZ}(\vec{r}) + H(\vec{r})\frac{F_{AZ}(\vec{r})}{\bar{n}}, \qquad (3)$$

where $F_{AZ}(\vec{r})$ - density of the $\tilde{\nu}_e$-flux from the active zone, $H(\vec{r})$ - the exact value of the generalized hardness in the point $\vec{r}$.

As the cross section is the additive value then (similar to (2) and (3)) for the inverse beta decay reaction ($\bar{\nu}_e+p \rightarrow n+e^+$) we can write the cross section for the total $\tilde{\nu}_e$-spectrum:

$$\sigma_{\bar{\nu}_e p}(\vec{r}) = \sigma_{\bar{\nu}_e p}^{AZ} + H(\vec{r}) \times \sigma_{\bar{\nu}_e p}^{Li}. \qquad (4)$$

The threshold of the reaction is 1.8 MeV but often (depending on the background) the used threshold is 3 MeV. Taking into account the data of [5] the cross section (4) was calculated as function of the hardness $H$ for the $E_{threshold} = 3$ MeV (see Fig. 3a (the left figure )).

At increase of $H$-value the double rise (and more) of the cross section is caused by enlarged part of lithium neutrinos in the total spectrum and energy squared dependence $\sigma_\nu \sim E_\nu^2$. For lithium spectrum the relative yield to the cross section (4) ensured by more high energy neutrinos is significantly larger compare to the the reactor spectrum (here in calculation we used $\bar{\nu}_e$-spectrum of $^{235}$U [6] as a single fuel isotope). This fact suggest to us to recalculate the cross section for more higher thresholds. The results (for thresholds - $E_{threshold} = 4, 5$ and 6 MeV) show that for hard total spectrum the lithium yield to the cross section strongly dominates the reactor part (Fig. 3a (the left figure )).

The partial yield of lithium antineutrinos increases with growing of the hardness of the total $\bar{\nu}_e$-spectrum. But in spite of decrease of lithium $\bar{\nu}_e$-flux (per fission) with increase of thresholds the normalized yields of lithium antineutrinos to the cross section of ($\bar{\nu}_e+p \rightarrow n+e^+$)-reaction (in the total spectrum) will rise more rapidly (as function of $H$) due to the low $\sigma_\nu \sim E_\nu^2$. It it possible to illustrate this: if to normalize the second summand in the equation (4) on the weight $W_{E_{\bar{\nu}} > E_{threshold}}$ of lithium antineutrinos in it's spectrum above the thresholds $E_{\bar{\nu}_e} > E_{threshold}$ (i.e., on 0.916, 0.828, 0.711 and 0.575 for 3, 4, 5 and 6 MeV, correspondingly), then the cross section functional ($\sigma_{\bar{\nu}_e p}^{AZ} + H(\vec{r}) \times \sigma_{\bar{\nu}_e p}^{Li} / W_{E_{\bar{\nu}} > E_{threshold}}$) for higher thresholds will rise more rapidly as function of hardness $H$. As a result the functional values come closer to each other for larger hardness $H$ ( see Fig. 3b (the right figure )).

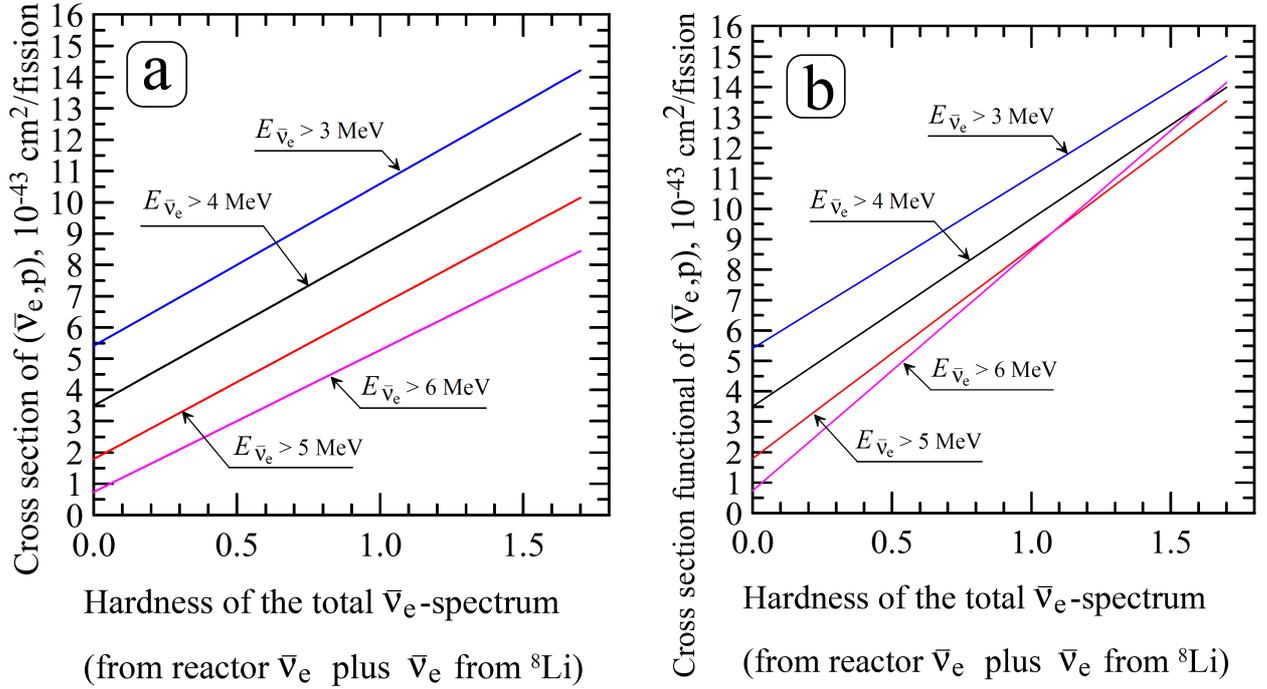

Fig. 3. On the left (Fig. 3a): cross section of ($\bar{\nu}_e + p \to n + e^+$)-reaction in the total $\bar{\nu}_e$-spectrum as function of the hardness $H$. The results are given for different thresholds of registration: 3, 4, 5 and 6 MeV. On the right (Fig.3b): cross section functional of ($\bar{\nu}_e + p \to n + e^+$)-reaction in the total $\bar{\nu}_e$-spectrum as function of the hardness $H$ (for the same thresholds).

The reactor $\bar{\nu}_e$-spectrum is known with significant errors (for all four main fuel isotopes $^{235}$U, $^{238}$U, $^{239}$Pu, $^{241}$Pu) which strongly rise with increase of the energy. Fig. 4 demonstrates the large difficulties in adequate specification of the $^{235}$U spectrum for high energy part.

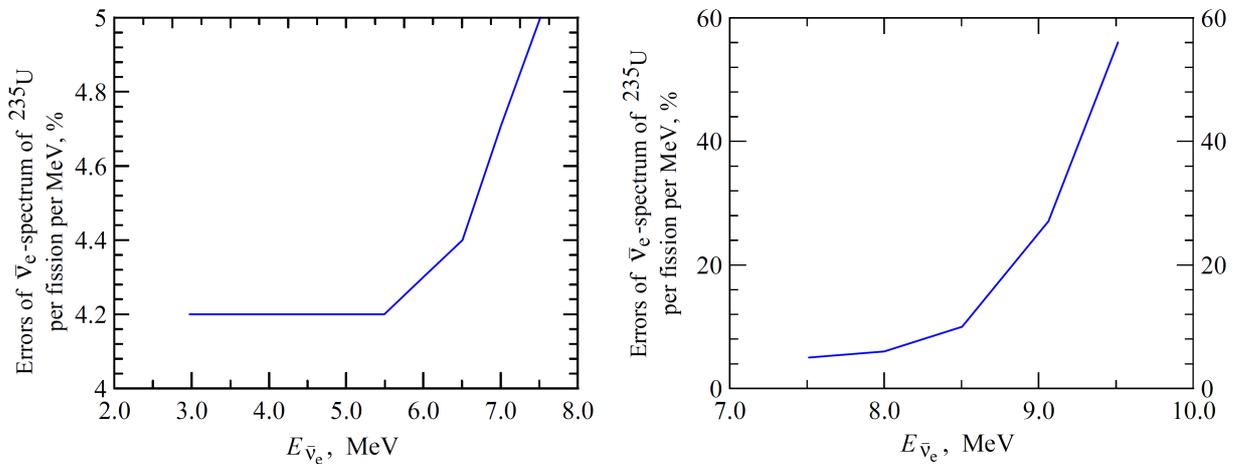

Fig. 4. Errors of the pure $\bar{\nu}_e$-spectrum of $^{235}$U (i.e., hardness $H = 0$): for $E_{\bar{\nu}} < 7.5$ MeV - see left plot; for $E_{\bar{\nu}} > 7.5$ MeV - see right plot. The results are presented for the $E_{threshold} = 3$ MeV.

In order to evaluate the advantages (in error values) given by well known $^8$Li spectrum we calculated the dependence of errors (in the total $\bar{\nu}_e$-spectrum) on the energy of antineutrinos (see Fig. 5 for the threshold 3 MeV). The family of curves demonstrate the rapid drop of spectrum errors for increasing energy and hardness $H$.

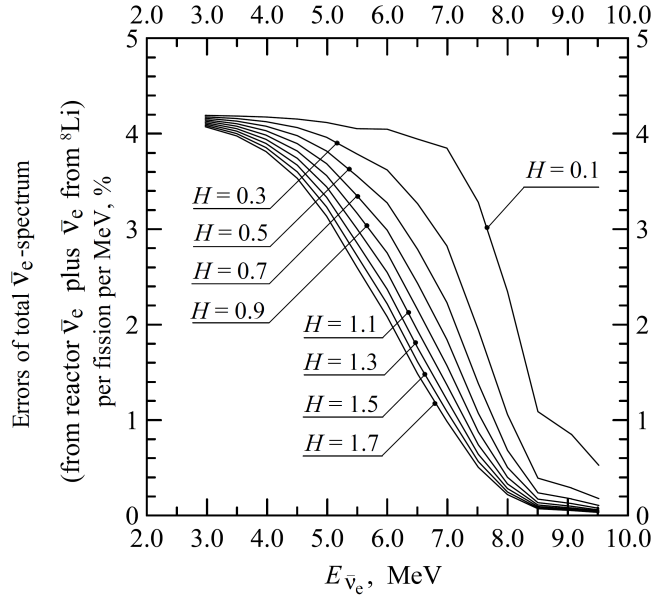

Figure 5. Errors of the total $\bar{\nu}_e$-spectrum for different hardness $H$. The all results are presented for the $E_{threshold} = 3$ MeV.

Then these errors were averaged on their total $\bar{\nu}_e$-spectra for every of thresholds: $E_{threshold} = 3, 4, 5, 6$ MeV. The final dependences of averaged errors on hardness $H$ for specified thresholds are presented in the Fig. 6.

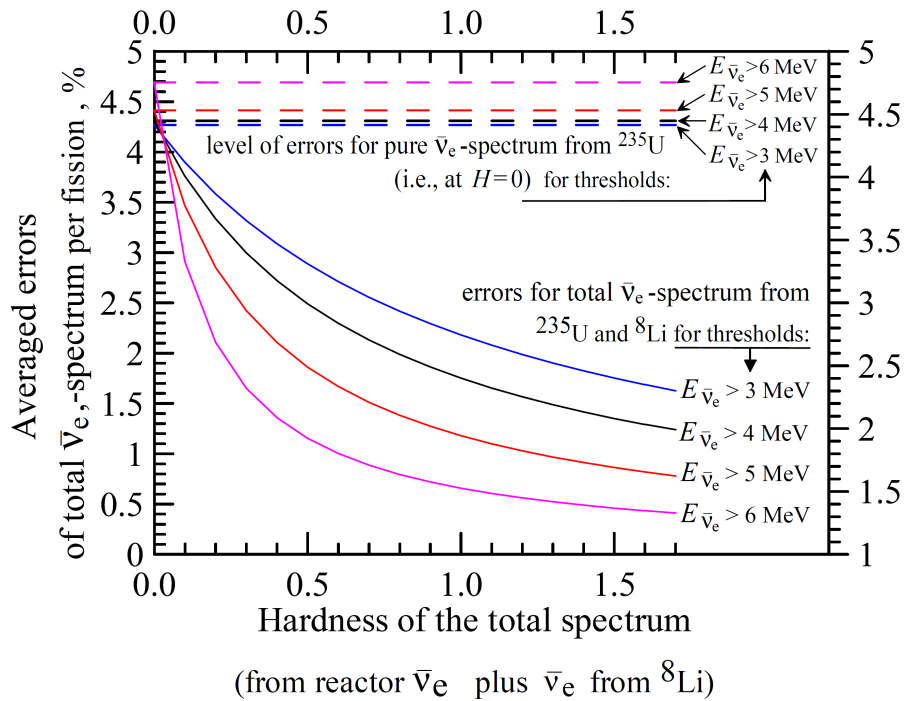

Fig. 6. Dependence of the averaged errors of the total $\bar{\nu}_e$-spectra from value of the hardness $H$ for the indicated $E_{threshold} = 3, 4, 5$ and 6 MeV.

The plot gives us the graphic solution: if you want to decrease the errors to the desired value (%) then you need to increase the hardness or to enlarge the threshold. So, for example (see Fig. 6), for 2.5% -error level the task will be solved if to realize the next every pairs of values (hardness $H$ and the corresponding threshold): $H=0.75$ for the threshold 3 MeV; $H=0.5$ - for 4 MeV; $H=0.28$ - for 5 MeV; $H=0.15$ for 6 MeV. For considered example the rate of detected ($\bar{\nu}_e+p$) -events will be in relationship 4.6 : 3.0 : 1.6 : 1.0, correspondingly. So, the more preferable way to minimize the spectrum errors is to enlarge the hardness as the alternative action (to increase the threshold) will cause drop in the detected events. Thus the proposed scheme of the $\bar{\nu}_e$ -source with variable spectrum allows to vary the hardness of the total $\bar{\nu}_e$ - spectrum for the purposes of precision .

## 3. THE POSSIBLE SETUP FOR SEARCH OF STERILE NEUTRINOS

The lithium antineutrino spectrum is attractive for oscillation experiments owing to well known distribution and it's hardness. These spectrum features can be especially helpful for search of sterile neutrinos. The problem of mass scale $\Delta m^2$ (between sterile and active neutrinos) is discussed intensively and some results indicate on eV-scale [7]. In this case the length of oscillation for $^8$Li antineutrinos will be about ~10 m and search of oscillations must be concentrated on short base line experiments [8,9]. Regimes with very fast lithium delivery and long distance from the blanket to the pumped volume (reservoir) can ensure large values of hardness $H$ not far from the reservoir. In such installations the advantages of the hard lithium spectrum ($\bar{E}_{\bar{\nu}}$ =6.5 Mev) will "answer" the demands of short base experiments. The real experimental geometry may be similar to configuration as in Fig. 7.

For oscillation discovery the position of the detector must correspond to the maximal oscillation signature of the checked model. The oscillation phase (of $\bar{\nu}_e$ escaped from the reservoir and blanket as the main lithium volumes) in the detector position must be equal and the detector must be settled as possible nearer to the pumped reservoir (in order to ensure the more hard $\bar{\nu}_e$ -spectrum). So, for the scheme (3+2) with two sterile neutrinos and matrix elements $\Delta m^2_{41} = 0.47$ eV$^2$, $U_{e4} = 0.128$, $\Delta m^2_{51} = 0.87$ eV$^2$, $U_{e5} = 0.138$ [7] the possible distance for search of sterile neutrinos can be $L_2$ = 3-6 m (as in this scheme the maximal oscillation signature ~0.89 appears at $L_2$ = 5 m). The variant of the neutrino detector construction IND used at Rovno nuclear power station is presented in [10].

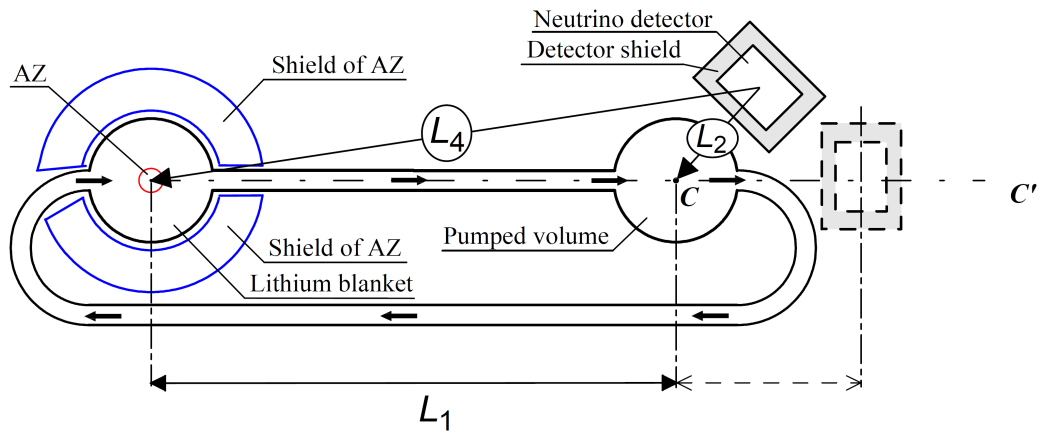

Figure 7. The scheme of the installation and possible position of $\bar{\nu}_e$ -detector. $L_2$ and $L_4$ - the remoteness of the detector from the center of the pumped reservoir and lithium blanket around the active zone (AZ) correspondingly. The preliminary discussed position of the neutrino detector (see Fig. 1) is shown by dotted lines. The identifications $L_1$, $L_2$ and $L_4$ are the same as in Fig. 2 of [1].

## 4. CONCLUSION

Today none of any artificial neutrino sources (except for nuclear explosions) can not able to produce such high flux of lower energy neutrinos as reactor. But significant errors of it's spectrum strongly complicate understanding of the results for oscillation experiments. The attractive solution will be to use the reactor as an intensive neutron source for (n,γ)-activation of $^7$Li and in so way to create a high flux antineutrino with large hardness of the total $\bar{\nu}_e$-spectrum (i.e., large part of lithium $\bar{\nu}_e$-flux in the resulting antineutrino flux). Owing to the large hardness of the created total spectrum the errors will drop significantly especially for the more high energy part of the total $\bar{\nu}_e$-spectrum. Such approach will help to solve the problem of large uncertainties in interpretation of the results.

The work devoted to development of an intense antineutrino source with hard spectrum on the base of short lived $^8$Li isotope in the scheme of the continuous loop circulations. Realization of the fast delivery of the (n,γ)-activated lithium from the reactor active zone (as a source of neutrons) to the remote neutrino detector will ensure large hardness of the total $\bar{\nu}_e$-spectrum (from active zone and from $\beta^-$-decaying $^8$Li isotope) close to the neutrino detector. The scheme allows to change the velocity of lithium delivery (changing the rate of lithium transport in the channels) and in so way to vary the total spectrum hardness near the detector. This unique possibility allows to vary the hardness smoothly and without any changes in the configuration of the installation and interruption of the experiment.

It was considered the variants for considerable decrease of errors of the total $\bar{\nu}_e$-spectrum (from the reactor active zone plus from the decay of $^8$Li isotope) by rising the spectrum hardness. This analyses was made for different thresholds of $\bar{\nu}_e$-detection. The obtained results confirmed that owing to the hard and well defined lithium $\tilde{\nu}_e$-spectrum the errors of the total $\tilde{\nu}_e$-spectrum can be diminished in two times and more. These results clearly show that the obtained decrease in errors can help to solve the problem of large uncertainties in spectrum. So, averaged errors of the $\bar{\nu}_e$-spectrum for thresholds of $\tilde{\nu}_e$-registration 3, 4, 5 and 6 MeV are (see Fig. 6): for pure $\bar{\nu}_e$-spectrum of $^{235}$U - 4.27, 4.31, 4.41 and 4.96%; for total $\bar{\nu}_e$-spectrum at hardness H=1.0 - 2.18, 1.75, 1.18 and 0.66%. I.e., the errors for total $\bar{\nu}_e$-spectrum at hardness $H$=1.0 decrease in 1.96, 2.46, 3.74 and 7.52 times. for thresholds 3, 4, 5 and 6 MeV.

The results of errors analysis is important for precision of short base line oscillation experiments for search of sterile neutrinos with $\Delta m^2$ in the range about 1 eV$^2$ on which indicates the results of global fits of some oscillation experiments. The possible configuration of the installation with regulated and controlled total $\tilde{\nu}_e$-spectrum for search of sterile neutrino is discussed.

**ACKNOWLEDGEMENTS**
The author thanks Yu. S. Lutostansky for helpful and useful discussion. The author tender thanks to L. B. Bezrukov, B. K. Lubsandorzhiev and I. I. Tkachev for their interest and support of the work.